\documentclass[aps,prl,twocolumn,showpacs,groupedaddress]{revtex4}
\usepackage{epsfig}
\usepackage{dcolumn}   
\usepackage{bm}        
\usepackage{amssymb}

\begin{document}
\title{Di-Jet Conical Correlations Associated with Heavy Quark Jets\\ 
in anti--de Sitter Space/Conformal Field Theory Correspondence}

\author{Jorge Noronha$^1$, Miklos Gyulassy$^{1,2,3}$, and Giorgio Torrieri$^2$}
\affiliation{$^1$Department of
Physics, Columbia University, 538 West 120$^{th}$ Street, New York,
NY 10027, USA\\
$^2$Institut f\"ur Theoretische Physik, Goethe Universit\"at, Frankfurt, Germany\\
$^3$Frankfurt Institute for Advanced Studies, Goethe Universit\"at, Frankfurt, Germany}

\begin{abstract}

We show that far zone Mach and diffusion wake ``holograms'' produced by supersonic strings in anti--de Sitter space/conformal field theory (AdS/CFT)  correspondence do not lead to observable conical angular correlations in the strict $N_c\to\infty$ supergravity limit if Cooper-Frye hadronization is assumed. However, a special {\em nonequilibrium}  ``neck'' zone near the jet is shown to produce an apparent sonic boom azimuthal angle distribution that is roughly independent of the heavy quark's velocity. Our results indicate that a measurement of the dependence of the away-side correlations on the velocity of associated identified heavy quark jets at the BNL Relativistic Heavy Ion Collider and CERN LHC will provide a direct test of the nonperturbative dynamics involved in the coupling between jets and the strongly-coupled Quark-Gluon Plasma (sQGP) implied by AdS/CFT correspondence.
\end{abstract}


\date{\today}
\maketitle

The reported observation of Mach cone-like correlations between tagged (near side) jets fragments and away-side associated moderate $p_T<4$ GeV/c hadrons \cite{Adler:2005ee} has generated interest because it may provide further evidence \cite{Stoecker:2004qu,shuryakcone} for fast relaxation times and the near perfect fluid property of the sQGP produced in Au+Au reactions at RHIC \cite{Gyulassy:2004zy}. While perturbative quantum chromodynamics (pQCD) correctly predicted the differential suppression of light quark and gluon jets \cite{pqcdjets}, recent data on non-photonic single electrons (from heavy quark jet fragment decay) \cite{Adler:2005xv} have challenged pQCD interpretations of jet quenching phenomena \cite{Djordjevic:2005db}. The combined observations of high suppression of heavy quark jets, large elliptic anisotropy of both light and heavy quark jet fragments, and the Mach-like away-side correlations have motivated novel approaches to try to explain these phenomena in terms of the Anti-de Sitter/Conformal Field Theory (AdS/CFT) correspondence \cite{maldacena}. We shall show bellow that AdS/CFT models do predict novel phenomena especially in the heavy quark jet sector where future decisive experiments can be used to verify this new geometric modeling of strongly-coupled gauge theory physics.
\begin{figure}[tb]
\centering
%
\epsfig{file=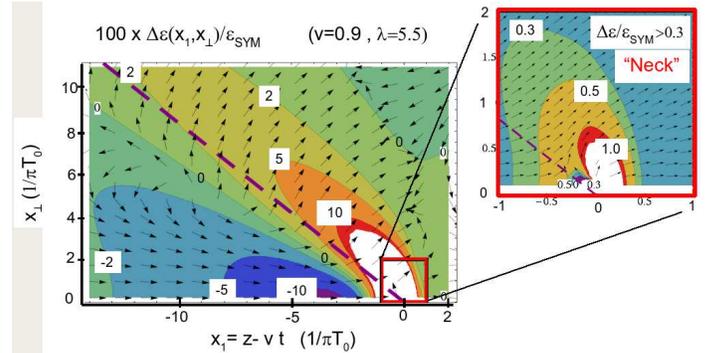,width=3.6in,clip=}
%
\caption{\label{angulardistplot}
(Color online) Energy density perturbation $\Delta\varepsilon(x_1, x_\perp)/\varepsilon_{SYM}$
from Ref.\ {\protect\cite{gubsermach}} due to a heavy quark jet with $v/c=0.9$
in a $\mathcal{N}=4$ SYM plasma modeled via the AdS/CFT string drag model for $N_c=3,\;\lambda=g^2_{YM}N_c=5.5$. Left panel shows the far zone (the numbers in the plot label the contours, in per cent as defined on the upper-left corner). The Mach wake zone is above the dashed line, $\cos\phi_M=1/(\surd 3 v)$, and the diffusion zone lies below that line.
Normalized Poynting (momentum flux) vector flow directions are indicated by arrows. The insert shows the nonequilibrium ``Neck'' zone (with the Coulomb Head subtracted) as defined by the condition that $\Delta\varepsilon/\varepsilon_{SYM}>0.3$.
Note the strong non-hydrodynamic transverse energy flow near the core.}
\end{figure}

The classical Nambu-Goto string drag solutions in an AdS$_5$ black brane background were proposed (see \cite{Gubser:2006bz}-\cite{Gubser:2008vz}, \cite{Herzog:2006gh}-\cite{Chesler:2007sv}, and \cite{Lin:2006rf} and refs. therein) as a detailed dynamical holographic description of the possible response of a strongly coupled 3+1D  Supersymmetric Yang Mills (SYM) plasma (an analog of sQGP) to the energy loss of a heavy quark jet modeled by the endpoint of a supersonic string. Even though supergravity ($N_c\rightarrow \infty$ and $\lambda=g^2_{YM} N_c\rightarrow \infty$) descriptions do not lead to true pQCD-like jet structures \cite{spherical}, the AdS$_5$ black brane string calculus provides a rare solvable example of how at least one strongly coupled gauge system could react to both subsonic and supersonic disturbances. In particular, these solutions feature both the familiar and universal far zone near equilibrium Mach sonic boom and diffusion wave collective flow phenomena shown in Fig.\ 1 as well as particular model dependent ``Neck'' and ``Head'' zones corresponding to the non-equilibrium dynamics involving coupled non-Abelian fields and hydrodynamic effects. Associated Mach-like conical angular correlations in nuclear collisions have been mainly studied in the past assuming near equilibrium hydrodynamic models \cite{Stoecker:2004qu,shuryakcone} and also pQCD jet energy loss models \cite{Vitev:2005yg}. In this letter we study the different sources of associated correlations taking into account the Mach wake and diffusion zones \cite{shuryakcone,gubsermach,Gubser:2007zr,Gubser:2008vz} as well as the out-of-equilibrium Neck zone \cite{Yarom:2007ni,Gubser:2007nd} shown in the insert of Fig.\ 1. General arguments based on the large $N_c$ limit are used to demonstrate that the yield from the hydrodynamic zone does not produce a conical pattern. The roughly velocity independent, strong conical flow from the Neck region is a true prediction of the AdS/CFT string drag model that can be verified in the near future at RHIC and LHC.

In the AdS/CFT trailing string scenario, the string disturbs the bulk 5-dimensional metric from which $T^{\mu\nu}(x)$ can be computed on the holographic Minkowski boundary of AdS$_5$
\cite{Friess:2006fk}. Because the heavy quark string is assumed to move at a constant speed, $v$, through the plasma there is an external source of stress, $\partial_{\mu} T^{\mu\nu}=J^{\nu}$ with $J^{\mu}(x)=J\,\delta(z-vt)\delta^2(\vec{x}_\perp)(v,1,\vec{0}_\perp)$ and $J=-dP/dt= \frac{\pi}{2}\sqrt{\lambda T_0^4 v^2/(1-v^2)}$ \cite{Friess:2006fk}, where $T_0$ is the temperature of the undisturbed background plasma. Beyond a certain ``speed limit'' \cite{Gubser:2006bz}, $p_T(max) \sim 4M_Q^3/(\lambda T_0^{2})$, where $M_Q$ is the heavy quark mass, the classical string drag picture breaks down due to the appearance of a horizon but at LHC there could be a sizable window  $p_T \stackrel{<}{\sim} 30$ GeV where it could be applied \cite{Horowitz:2007su}.

In the supergravity $N_c \rightarrow \infty$ limit,
the disturbances in the
energy-momentum tensor $\delta T^{\mu\nu}$ caused by the moving string are
of order $\sqrt{\lambda}/N_{c}^{2}$ with respect to
the background plasma energy density, $\varepsilon_{SYM}=K T_0^4$, with $K\propto N_c^2$ \cite{Gubser:1996de}.
Therefore, the disturbances are arbitrarily small in this limit
 because $\sqrt{\lambda}/N_{c}^{2}\to 0$.
An immediate consequence of this is that linearized
first-order Navier-Stokes hydrodynamics is predicted
to be a very good description
of the quark's holographic wake down to short distance scales away from the
heavy quark \cite{Chesler:2007sv,Noronha:2007xe,Gubser:2008vz}. In general, the flow velocity $U^{\mu}(X)=\left(\sqrt{1+\vec{U}^2},\vec{U}\right)$ can be obtained from $T^{\mu\nu}(X)$ by boosting the system to its local Landau rest
frame. We emphasize that the Coulomb stress contribution has been subtracted as in \cite{gubsermach}. When $N_c \rightarrow \infty$ the flow velocity reduces to $U^{i}=T^{0i}/\left(4 P_0\right)$,
where $P_0=\varepsilon_{SYM}/3 \propto N_c^2 T_0^4 $
is the background $\mathcal{N}=4$ SYM plasma pressure
 \cite{Chesler:2007sv,Noronha:2007xe}. Since
$T^{0i}$ is only proportional to $\sqrt{\lambda}$, we see that
$U\sim \mathcal{O}(\sqrt{\lambda}/N_{c}^2)$ is also tiny
in the strict supergravity
limit. Within this approximation, the local temperature is $T(X)=\left[8\, T^{00}(X)/3\pi^2 (N_c^2-1)\right]^{1/4}$. Thus, the local temperature fluctuates as $T(X)=T_{0}+\Delta T(X)$ with  $\Delta T(X)/T_0=
\mathcal{O}(\sqrt{\lambda}/N_{c}^2)$. On the other hand, in the Head and Neck zones
the stress grows rapidly and the stress perturbation
becomes comparable to the background. Since the Head zone is defined
by a large amplitude Lorentz contracted Coulomb field \cite{Friess:2006fk}, the near perfect fluid
hydrodynamic approximation must break down close to the quark.
In Ref.\ \cite{Noronha:2007xe} the analytic near field Yarom solution, $T^{\mu\nu}_{Y}(X)$ \cite{Yarom:2007ni}, was used to compute a ``Knudsen'' boundary defined by the closed surface inside which the effective Knudsen number $Kn \equiv\Gamma_s \left|\vec{\nabla}\cdot \vec{S}_Y\right|/|\vec{S}_Y|=1$. Here $\Gamma_s=1/\left(3\pi T_0\right)$ is the sound attenuation length and $S^i_Y=-T_{0i}^{Y}$ is the Yarom momentum flux. Inside this boundary local equilibrium cannot be maintained even by uncertainty principle limited thermalization rates \cite{Danielewicz:1984ww,Policastro:2001yc}.

For $v=0.99$, we found
$Kn \geq 1$ in a
region $|x_1\equiv z-vt|<2/\pi T_0$ and $x_\perp$ extending into the far zone
studies in \cite{Chesler:2007sv}. Parametrically,
the Knudsen  zone is defined  by
\begin{equation}
T^{\mu\nu}_{Kn} \equiv \theta(Kn(x)-1)\; T^{\mu\nu}_{Y}(x)
\sim \frac{\surd \lambda T_0^2\; \zeta^{\mu\nu}}{x_\perp^2+ \gamma^2 x_1^2}
\end{equation}
where $\gamma^{-2}=1-v^2$ and $\zeta^{\mu\nu}(x)$ is a dimensionless
angular function inside the boundary.
Even Navier-Stokes viscous hydrodynamics is inapplicable in this zone. Because $T^{\mu\nu}_{Kn}\sim \surd\lambda T_0^2/|x|^2$
we see that this zone can be considered
as a dynamic interference ``Neck'' zone
between the near equilibrium $\sim T_0^4$ plasma stress dominated
far zone and a near Head zone
dominated by the stress of the
Lorentz boosted $\sim \sqrt{\lambda}/|x|^4$ Coulomb field of the heavy quark.
The AdS/CFT analogs of electro-magneto-hydrodynamic effects play the dominant role in this Neck zone. Within the Neck zone, there is also an inner core Head zone where the stress becomes dominated by the
external Coulomb field of the quark.
The Head zone can therefore be defined as in
Ref. \cite{Dominguez:2008vd} by equating the analytic
Coulomb energy density \cite{Friess:2006fk,gubsermach}, $\varepsilon_C(x_{1},x_{\perp})$,
to the analytic near zone Yarom energy density
\cite{Yarom:2007ni}, $\varepsilon_Y(x_{1},x_{\perp})$.
This Coulomb head boundary is approximately given by
\begin{equation}
x_{\perp}^2 + \gamma^2 x_{1}^2 =
\frac{1}{(\pi T_0)^4}\frac{(2x_{\perp}^2+x_{1}^2)^2}{\gamma^4 x_{1}^2(x_{\perp}^2/2 + \gamma^2 x_{1}^2)^2}.
\end{equation}
The Head zone is thus a Lorentz contracted
surface with a maximal longitudinal thickness
near $ \Delta x_{\perp}\,\pi\, T_0 \sim 1/\gamma^{1/2}$
and $\Delta x_{1,C} \,\pi\, T_0 \sim 1/\gamma^{3/2}$ in agreement with
Ref.\ \cite{Dominguez:2008vd}. The numerical results found in \cite{Noronha:2007xe}  indicated
that for $v=0.99$ jets, the nonequilibrium dynamic screening Knudsen Neck zone is much thicker than the contracted Head zone, in agreement
with the parametric dependence $\Delta x_{1}^{Kn}\sim 2/\left(\pi T_0\right)  \sim 6\Gamma_s$ expected from uncertainty principle bounded dissipation rates \cite{Danielewicz:1984ww,Policastro:2001yc}.

The weakest link between any strongly coupled description and experimental data is the freeze-out:  The need to convert the strongly interacting fluid into weakly interacting (and eventually free) particles.  While the {\em dynamics} of how this happens in nature is very poorly understood from first principles, it should be noted that, as long as we use momentum observables and the hadronic rescattering is reasonably short, momentum conservation at freeze-out should limit the effect it can have on early time observables. The usual ansatz used is the Cooper-Frye (CF) formula, where the conversion of the fluid into free particles is achieved instantaneously at a critical surface $d \Sigma_\mu$ \cite{Cooper:1974mv}. Assuming such a freeze-out scheme \cite{shuryakcone,heinzcone}, we can obtain particle distributions and correlations from the flow velocity field $U^{\mu}(X)$ and temperature $T(X)$.

For associated (massless) particles with
$P^{\mu}=(p_{T},p_{T}\cos (\pi-\phi),p_{T}\sin
(\pi-\phi),0)$
the momentum distribution at mid rapidity
$y=0$ is
\begin{equation}
\frac{dN}{p_Tdp_Tdy d\phi}\Big
|_{y=0}=\int_{\Sigma_T}d\Sigma_{\mu}P^{\mu}\left[f(U^{\mu},P^{\mu},T)-f_{eq}\right]
\label{cooperfrye}
\end{equation}
where $p_T$ is the transverse momentum, $\Sigma(X)$ is the freeze-out hypersurface, and $f_0=\exp(-U^{\mu}P_{\mu}/T(X))$ is a local Boltzmann equilibrium distribution
that is approximately valid in the near equilibrium ``far'' zone $\Sigma_T=V\theta(1-Kn)$. We subtract  the isotropic SYM background yield via $f_{eq}\equiv f|_{U^{\mu}=0,T=T_0}$. Viscous corrections to the Boltzmann distribution function \cite{Dusling:2007gi} produce subleading contributions in $1/N_c$ that are negligible in the supergravity limit. Since the medium is static and infinite we use the isochronous ansatz $d\Sigma^\mu= x_\perp d x_\perp dx_1 d\varphi\,\left(1,0,0,0\right)$. More realistic prescriptions such as the isothermal freeze-out are more adequate and applicable when considering expanding media.

In the large $N_c\rightarrow \infty $ limit,  $\Delta T \ll T_0$ and $\vec{p}\cdot\vec{U}\ll T_0$ as noted before and the Boltzmann exponent can be expanded
up to corrections $\mathcal{O} (\lambda/ N_{c}^4)$. There is axial symmetry with respect to the trigger jet axis defined here as $n^\mu\equiv(0,1,0,0)$, where the ``$x_1=z-vt$'' axis of the away-side jet corresponds to $-n^\mu$ and the nuclear beam axis corresponds to the ``y'' axis. In these coordinates, $U^{\mu}(x_1,x_\perp)= ( U^0, U_1, U_{\perp}\sin\varphi, U_\perp \cos\varphi)$. The associated away-side azimuthal distribution at mid-rapidity $f(\phi)=dN/p_{T}dp_{T}dyd\phi |_{y=0}$ with respect to the beam axis is then given after integrating over $\varphi$ by
\begin{eqnarray}
f(\phi)=2\pi\,p_T\,\int_{\Sigma_{T}} dx_1 dx_\perp x_\perp \times
\label{fulldistrib}
\end{eqnarray}
\[\ \left(
\exp\left\{-\frac{p_T}{T}\left[U_0-U_1\cos(\pi-\phi)\right]\right\}\, I_0(a_\perp)-e^{-p_T/T_0} \right) \]
where $a_\perp=p_\perp U_\perp\sin(\pi-\phi)/T$ and $I_0$ is the modified Bessel function. In the supergravity approximation $a_{\perp} \sim \mathcal{O}\left(\frac{\sqrt\lambda}{N_{c}^2}\right) \ll 1$ and, thus, we can expand the Bessel function $\lim_{x\to 0}\,I_0 (x) =1+\frac{x^2}{4}+\mathcal{O}(x^4)$ to get the approximate equation for the distribution
\begin{equation}
f(\phi) \simeq e^{-p_{T}/T_0}\frac{2\pi\,p_{T}^2}{T_0}
\left[\frac{\langle \Delta T\rangle}{T_0} + \langle U_{1}\rangle \cos
(\pi-\phi) 
\right]\label{expandedCFfinal}
\end{equation}
where deviations from isotropy are then controlled by the following global moments $\langle \Delta T\rangle= \int_{\Sigma_T} dx_1 dx_\perp x_\perp \,\Delta T$ and $\langle U_1\rangle= \int_{\Sigma_T} dx_1 dx_\perp x_\perp \,U_1$.

It is clear that in the strict $N_c\gg 1 $ limit the azimuthal distribution only has a trivial broad peak at $\phi=\pi$. A double-peaked structure in the away-side of the jet correlation function can only arise from the Mach zone when $N_c$ is reduced so that $I_0$ in Eq.\ (\ref{fulldistrib}) deviates from unity (see Ref.\ \cite{shuryakcone} where it was shown that the angular correlations associated with very soft particles do not have a conical structure within linearized hydrodynamics).

When $N_c=3$ the simplifications due to the supergravity approximation are not strictly valid but it is of interest to use the numerical solutions for $T^{0\mu}$ computed by Gubser, Pufu, and Yarom (GPY) \cite{gubsermach} to check how large the induced stress can be for ``realistic'' parameters. As seen in Fig.\ 1, even for $\lambda=5.5, N_c=3$ the deviations remain less than 10\% outside the Neck region. We chose our CF volume to be defined by the forward light-cone that begins at $-14/(\pi T_0)< x_1<2/(\pi T_0)$ and $x_\perp< 14/(\pi T_0)$, which corresponds to a cylinder of roughly $L\sim R\sim 5$ fm at $T_0=0.2$ GeV. Note that $T(X)$ is not well defined in a small region within the Head zone. In practice, to avoid this issue we took $x_{\perp} \pi T_0 > 0.2$ when computing the hadronic yield. Note, however, that we do not consider the Head yield in the following. We take $p_T=4-7\, \pi T_0\sim 2.5-4.4$ GeV for the associated hadron as a typical momentum of interest at RHIC. The Neck zone was defined by the condition $\Delta \varepsilon/\varepsilon_{SYM} >0.3$ corresponding to roughly the small square in Fig.\ 1 \cite{Noronha:2008tg}. The far zone minus the Neck zone gave rise to the blue azimuthal distributions in Fig.\ 2. There, three jet velocities, $v=0.9,\,0.75,\,0.58$ are shown where $p_T=4$-$5$, $5$-$6$, $6$-$7\, \pi T_0$, respectively. In all cases the far zone distribution peaks at $\phi=\pi$ even for these more ``realistic'' parameters. We have also checked that no Mach dip appears in that region even for much higher $p_T \sim 20\,\pi T_0$.
\begin{figure}[ht]
\centering
%
\epsfig{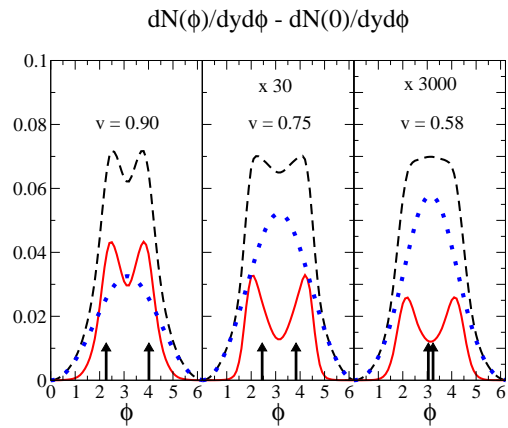}
\caption{\label{angulardistplot}
(Color online) Mid-rapidity azimuthal away-side associated angular
distribution from the Cooper-Frye freeze-out of the $(T(x),\vec{U}(x))$ fields extracted from \cite{gubsermach}, based on the AdS/CFT string drag model. Three cases for various heavy quark jet velocity and associated hadron transverse momentum ranges,
$1: (v/c=0.9, p_T/\pi T_0=4$-$5$), $2: (v/c=0.75, p_T/\pi T_0=5$-$6$), and $3: (v/c=0.58, p_T/\pi T_0=6$-$7$), are compared. Note the scale factors in the plots. The short arrows show the
expected Mach angles. The yields from the Neck region (solid red), Mach and diffusion zones (dotted blue), and the sum from all contributions (dashed black) are shown in this plot.}
\end{figure}
An application of CF freeze-out without linearization (i.e. Eq. (\ref{fulldistrib}))  to the Neck zone leads to red curves that exhibit a clear double shoulder structure remarkably close to the expected Mach angles for  $v=0.9 $. The total dashed black distribution could be easily misinterpreted in this case as a signature for Mach cone emission. However, this angular structure is shown to be almost independent on the jet velocity. We have verified that curves from the Neck region also follow if the numerical GPY solution is replaced by the analytic Yarom solution in the Neck zone. While CF may not be a reliable hadronization scheme for the Neck zone, it correctly shows the presence of the strong transverse Poynting vector flow field seen in the insert of Fig.\ 1.  The conservation of energy flux implies that this correlation may be robust to more general hadronization schemes as we will show elsewhere. In general, the underlying flow of the medium is expected to modify the effective Mach cone angle \cite{heinzcone,Satarov:2005mv}. The preliminary experimental finding \cite{Winter:2008bn} that the location of the experimental peaks does not change with either centrality or the orientation with respect to the reaction plane casts doubt that the origin of the measured associated conical correlations are due to ordinary Mach cones. A confirmation of these results may give support to the new source of conical correlations found in this paper for the case of associated heavy quark jets.

In conclusion, while AdS/CFT string solutions feature Mach wakes in coordinate space, their signal is too weak in supergravity to produce observable double shoulder correlations. However, the nonequilibrium  transverse flux from the Neck zone could imitate Mach cone-like correlations, without, however, the dependence of the angle on the jet velocity expected from Mach's law. The new results for azimuthal correlations from the near and far zones presented above are generic to AdS/CFT based modeling of heavy quark jets. We propose that a measurement of the jet velocity dependence of the away-side distributions associated with future tagged heavy quark jets at RHIC and LHC can be used to look for the novel source of conical correlations discussed in this letter.

We thank S. Gubser, S. Pufu, and A. Yarom for extensive discussions and for providing tables of their numerical solutions and B.\ Betz, H.\ St\"ocker, D.\ Rischke, and C.\ Greiner for discussions. J.N. and M.G. acknowledge support from US-DOE Nuclear Science Grant No. DE-FG02-93ER40764 and M.G. is grateful for DFG Mercator Gast Professor support while on sabbatical at ITP/Goethe University. G.T. thanks the Alexander Von Humboldt foundation and Goethe University for support.

\end{document}